# Influence of Asymmetric Gain Suppression on Relative Intensity Noise Properties of Multimode Semiconductor Lasers

[1]Ahmed Bakry, [1]Ahmed Alshahrie, [2]Alaa Mahmoud, [3]Hamed Dalir and [1,*]Moustafa Ahmed

[1]Department of Physics, Faculty of Science, King Abdulaziz University, 80203 Jeddah 21589, Saudi Arabia

[2]Laser Institute for Research and Applications (LIRA), Beni-Suef University, Egypt.

[2]Department of Electrical and Computer Engineering, George Washington University, Washington, DC 20052, USA

[*]Corresponding author: mostafa.farghal@mu.edu.eg

**Abstract**

We introduce modeling and simulation of the noise properties associated with types of modal oscillations induced by scaling the asymmetric gain suppression (AGS) in multimode semiconductor lasers. The study is based on numerical integration of a system of rate equations of 21-oscillating modes taking account of the self- and cross-modal gain suppression mechanisms. AGS is varied in terms of a pre-defined parameter, which is controlled by the linewidth enhancement factor and differential gain. Basing on intensive simulation of the mode dynamics, we present a mapping (AGS versus current) diagram of the possible types of modal oscillations. When the laser oscillation is hopping multimode oscillation (HMMO), the spectra of relative intensity noise (RIN) of the total output and hopping modes are characterized by a sharp peak around the relaxation oscillation (RO) frequency and a broad peak around the hopping frequency. The levels of RIN in the regimes of single-mode oscillation (SMO) are much lower than those under HMMO, and the mode-partition noise is two order of magnitudes lower.

# 1. Introduction

The current developments in research on the semiconductor laser speed have contributed effectively to a leap in the speeds of modern photonic laser applications, such as optical fiber links. Conversion of these lasers into a commercial form that brings photonics applications to the user end of fiber links, such as fiber to home the home (FTTH) networks, requires production of inexpensive high-speed lasers [39-44]. Also, reducing the level of noise in these lasers to the quantum or intrinsic level is a challenge in both research and technology.

Semiconductor lasers using Fabry-Perot (FP) resonators are known as low-cost radiation sources. Hoverer, even if the FP semiconductor laser is well controlled in the lateral and transverse directions, it happens to oscillate in several longitudinal modes [1,2]. In semiconductor lasers emitting in the C-band of fiber communications, termed as "long-wavelength lasers" such as InGaAsP lasers, the spectral gain is shallow and asymmetric around the lasing mode, which induces multi-longitudinal mode oscillation [3-7]. HMMO was observed in experiments and ascribed to large values of the linewidth enhancement factor that strengthens AGS and mode coupling [8]. Since each of these modes has its own intensity and phase noises, which is called "mode-partition noise (MPN)", both the intrinsic MPN and total noise could be increased if the oscillating modes couple in such a way to induce HMMO [9]. This degradation of the noise performance of the laser negatively affects its applications, for example MPN could be amplified when laser radiation propagates in dispersive optical fibers and increase the noise figure of the fiber link [1,10,11].

The multimode hopping could be released, and the laser supports SMO by subjecting the laser to very strong optical feedback and avoiding the chaotic state [12-14]. A cost-effective solution to suppress mode hopping and the induced MPN is to control the AGS and bias current [15]. Most recently, Mahmoud and Ahmed [15] have investigated the modal oscillations of FP semiconductor laser following a deterministic model of the multimode rate equations by scaling AGS and current. They predicted possible types of HMMO, symmetric multimode oscillation (SMMO), asymmetric steady-state multimode oscillation (ASMMO), and SMO.

In this paper, we extend our work in [15] to investigate the noise characteristics of the long-wavelength multimode semiconductor laser. The noise is quantified in terms of the frequency spectrum of RIN. We introduce characterization of the RIN spectrum in each of the investigated types of modal oscillations. The simulations in [15] were based on a deterministic form of the multimode rate equations that did not include the intrinsic intensity fluctuations of modes. However, it was shown in [7,16] that adding these noise sources could change the dynamics state of the laser, especially at their operating boundaries. Therefore, we re-investigate the operating regions of the modal oscillation types of the laser by adding Langevin noise sources to the multimode rate equations. AGS is varied in terms of a pre-defined parameter, which is controlled by the linewidth enhancement factor and differential gain. Basing on intensive simulation of the mode dynamics, we modify the mapping (AGS versus current) diagram of modal oscillations in [15]. We show that when the laser oscillation is HMMO, the RIN spectra of the total output and hopping modes are characterized by a sharp peak around the RO frequency as well as a broad peak around the hopping frequency. The levels of RIN in the regimes of SMO are much lower than those under HMMO. Also, MPN of the HMMO is two order of magnitudes higher than that under SMO.

The paper is organized as follows. In section 2, we introduce the theoretical stochastic rate equation model used in simulation. The numerical procedures are given in section 3, while the simulation results on the modal oscillations and noise properties are presented in section 4. Finally, conclusions appear in section 5.

## 2. Theoretical and calculation model of analysis

The dynamics and intensity fluctuations of the multi-longitudinal oscillating modes are described by the following stochastic version of the rate equations of the injected carrier number $N$ and the photon number $S_p$ of the oscillating modes [7]

$$\frac{dN}{dt} = \frac{I}{e} - \sum_p G_{L(p)} S_p - \frac{N}{\tau_s} + F_N(t) \tag{1}$$

$$\frac{dS_p}{dt} = [G_p - G_{th}] S_p + C_p \frac{N}{\tau_s} + F_{Sp}(t), \tag{2}$$

where $p = 0, \pm 1, \pm 2, \pm 3,\ldots$ is the mode number with $p = 0$ as the central mode of the gain spectrum of the active region. The wavelength of the other modes is $\lambda_p = \lambda_0 + p\Delta\lambda$ where $\Delta\lambda$ is the mode wavelength separation. The gain of mode $p$ per unit time is described as [7]:

$$G_p = A_p - BS_p - \sum_{q \neq p} \left[ D_{p(q)} + H_{p(q)} \right] S_q \tag{3}$$

where $A_p$ is called linear gain of mode $p$ whose spectrum is determined by the differential gain coefficient $a$ of the linear dependence of gain on the injected carrier number $N$ above the transparency level $N_g$, and the width $b$ of the parabolic spectrum as,

$$A_p = \frac{a\xi}{V}\left[ N - N_g - bV(\lambda_p - \lambda_{peak})^2 \right] \quad p = 0, \pm 1, \pm 2, \ldots \tag{4}$$

The central mode $p = 0$ is assumed to coincide with the peak wavelength of the gain spectrum. The other terms of Eq. (3) represent the nonlinear gain coefficient contributing to gain suppression. The coefficient $B$ is the self-modal suppression coefficient, and is determined by,

$$B = B_c(N - N_s) \tag{5}$$

where $B_c$ and $N_s$ are parameters of gain and carrier number characterizing this gain suppression. The last two terms describe modal suppression of gain of mode $p$ by other modes $q \neq p$. $D_{p(q)}$ and $H_{p(q)}$ are the coefficient of symmetric and asymmetric gain suppressions, respectively, and are given by [15,17]

$$D_{p(q)} = \frac{4}{3} \frac{B}{\left(2\pi c \tau_{in}/\lambda_p^2\right)(\lambda_p - \lambda_q)^2 + 1} \tag{6}$$

$$H_{p(q)} = H_c \frac{\lambda_p^2}{\lambda_q - \lambda_p}(N - N_g) \tag{7}$$

where

$$H_c = \frac{3\alpha}{8\pi c}\left(\frac{a\xi}{V}\right)^2 \tag{8}$$

is a parameter used to control the strength of AGS in this study. $c$ is the speed of light in free space and $\tau_{in}$ is the intraband relaxation time. As indicated in the above equation, $H_c$ is a function of both the differential gain coefficient $a$ and linewidth enhancement factor $\alpha$ [15]. Large values of $H_c$ indicate strong AGS, which works to enhance gain modes with longer wavelengths but suppress gain of shorter-wavelength modes. Inclusion of the spontaneous emission into the lasing mode $p$ is described in terms of the spontaneous emission factor $C_p$ in Eq. (2), which is given as [18]

$$C_p = \frac{a\xi\tau_s/V}{[2(\lambda_p - \lambda_0)/\delta\lambda]^2 + 1} \tag{9}$$

where $\delta\lambda$ is the half width of the spontaneous emission profile

The fluctuations on the carrier number $N$ and modal photon number $S_p$ are described by adding Langevin noise sources $F_N(t)$ and $F_{Sp}(t)$ to rate equations (1) and (2), respectively. These noise sources have Gaussian probability distributions with zero mean values, and are $\delta$-correlated as [7,19]

$$\langle F_a(t) F_b(t') \rangle = V_{ab}\delta(t - t') \tag{10}$$

where $a$ and $b$ stand for any of $N$ or $S_p$. The correlation variances $V_{ab}$ are determined from the steady state solutions of the rate equations at each integration step. The noise content of the fluctuating mode intensity is measured in terms of RIN using the time fluctuations $\delta S(t) = S(t) - \bar{S}$ in the total photon number $S(t) = \sum_p S_p(t)$ around its average value $\bar{S}$. Over a finite time, $T$, RIN is calculated from the Fourier transformation [20]

$$RIN = \frac{1}{\bar{S}^2}\left\{\frac{1}{T}\left|\int_0^T \delta S(t) e^{-j2\pi f\tau} d\tau\right|^2\right\} \tag{11}$$

where $f$ is the Fourier frequency. RIN of the individual modes is defined similarly in terms of the fluctuations of their photon number $S_p(t)$.

Definitions of the other laser parameters in the above equations and their typical values are given in Table 1.

**Table 1.** Definition and typical values of the parameters of the FP-InGaAsP laser used in the present calculations.

| Parameter | Meaning | Value | Unit |
|---|---|---|---|
| $\lambda_{peak}$ | Emission wavelength | 1.55 | Mm |
| $a$ | Deferential gain coefficient | $[1 - 10] \times 10^{-12}$ | $m^3s^{-1}$ |
| $\xi$ | Field confinement factor in the active layer | 0.2 | -- |
| $V$ | Volume of the active region | $150 \times 10^{-18}$ | $m^3$ |
| $N_g$ | Electron number at transparency | $1.33 \times 10^8$ | -- |
| $B$ | Dispersion parameter of the linear gain spectrum | $9.07 \times 10^{17}$ | $m^2A^{-2}$ |
| $\varepsilon_o$ | Electric permittivity | $8.85 \times 10^{-12}$ | $Fm^{-1}$ |
| $n_r$ | Refractive index of active region | 3.513 | -- |
| $\tau_{in}$ | Electron intraband relaxation time | 0.1 | ps |
| $|R_{cv}|^2$ | Squared absolute value of the dipole moment | $9.52 \times 10^{-57}$ | $C^2m^2$ |
| $N_s$ | Electron number characterizing gain suppression | $1.01 \times 10^8$ | -- |
| $\alpha$ | Linewidth enhancement factor | $[1 - 5]$ | -- |
| $\tau_s$ | Electron lifetime by spontaneous | 1 | ns |
| $\Delta\lambda$ | Half-width of spontaneous emission | 23 | nm |
| $G_{th}$ | Threshold gain level | $2.88 \times 10^{11}$ | $s^{-1}$ |
| $I_{th}$ | Threshold current | 25 | mA |

### 3. Numerical Calculations

The system of multimode rate equations (1) and (2) is integrated numerically using the fourth-order Runge–Kutta algorithm using a time step of integration as short as $\Delta t = 5$ ps to ensure good resolution of the time trajectories of the modal photon number $S_p(t)$. The integration is taken over a period of $T = 2$ μs, which is long enough to collect time fluctuations $\delta S_p(t)$ and to simulate a RIN spectrum with high resolution using the fast Fourier transform (FFT) as

$$RIN = \frac{1}{\bar{S}^2} \frac{\Delta t^2}{T} |FFT[\delta S(t_i)]|^2 \quad (12)$$

We assume oscillations of a large number of 21 modes. The output spectrum of the laser is calculated by constructing a bar plot of the time-averaged values $\bar{S}_p$ of the modal photon number versus the mode number $p$. This laser output spectrum determines the side mode suppression ratio (SMSR), defined as the ratio of the average photon numbers of the lasing mode to that of the strongest side mode. Values of SMSR > 100 corresponds to oscillation in single mode, SMO.

At each integration instant $t_i$, the noise sources $F_{Sp}(t_i)$ and $F_N(t_i)$ are generated using the following forms [7]

$$F_{Sp}(t_i) = \sqrt{\frac{V_{SpSp}(t_i)}{\Delta t}} g_{Sp} \quad (13)$$

$$F_N(t_i) = \sqrt{\frac{V_{NN}(t_i) - 2\sum_p V_{NSp}^2(t_i)/V_{SpSp}(t_i)}{\Delta t}} g_N + \sum_p \frac{V_{NSp(t_i)}}{V_{SpSp}(t_i)} F_{Sp}(t_i) \quad (14)$$

The variances $V_{ab}$ (with $a$ and $b$ referring to each of the symbols $N$ or $S_p$) at time $t_i$ are evaluated from $S(t_i - 1)$ and $N(t_i - 1)$ at the preceding time $t_i - 1$ by assuming quasi-steady states $(dS_p/dt \approx dN/dt \approx 0)$ over the integration step $\Delta t = t_{i-1} - t_i$ [20,21] as

$$V_{SpSp}(t_i) = 2\left[\frac{a\xi}{V} S_p(t_{i-1}) + \frac{C_p}{\tau_s}\right] N(t_{i-1}) \quad (15)$$

$$V_{NN}(t_i) = 2\left[\frac{1}{\tau_s} + \frac{a\xi}{V} \sum_p S_p(t_{i-1})\right] N(t_{i-1}) \quad (16)$$

$$V_{NSp}(t_i) = -\frac{a\xi}{V}\left[N(t_{i-1}) - N_g\right] S(t_{i-1}) - \frac{N(t_{i-1})}{\tau_s} \quad (17)$$

In Eqs. (13) and (14), $g_N$ and $g_{sp}$ are independent Gaussian random numbers with variances of unity and zero mean values. They are obtained at each integration step by applying the Box–Mueller approximation [22] to a set of uniformly distributed random numbers generated by the computer.

### 4. Simulation results and discussion
*4.1 Types of modal oscillation as functions of AGS*

The modal oscillations of the FP laser depend on competition among the oscillating longitudinal modes which is controlled by AGS. By dropping the noise sources from rate equation (1) and (2) and simulating the mode dynamics, Mahmoud and Ahmed [15] investigated the possible types of the modal oscillations of such a laser; namely,

MMHO: multimode hoping oscillations in which a number of modes lying on the long-wavelength side of the gain spectrum carries the total output and exhibit hoping in a semi-periodic fashion.

SMMO: symmetric multimode oscillations in which a number of modes lying on both sides of central gain carries most of the laser output.

ASMMO: asymmetric steady-state multimode oscillations in which the total output is contained a number of modes lying on the long-wavelength side of the gain spectrum and these modes reach a steady state.

SMO: single-mode oscillations in which the total output is mainly contained in one mode either the central mode or one with longer wavelength.

Since the noise sources may shift the boundaries of the dynamic states of the laser [16], we re-investigated the operating regions of the possible types of modal oscillations of the laser using the stochastic rate equations (1) and (2). The simulations are done as functions of the AGS parameter $H_c$ and current $I$. These results are then used to depict the ($H_c$ versus $I$) diagram of figure 1, which allocate the boundaries of each of these types. As shown in the figure, the laser operates in SSMMO near the threshold current, $I \leq 1.12 I_{th}$, regardless the value of $H_c$. ASMMO follows SMMO with the increase of current as long as $H_c < 0.125$ m$^{-1}$s$^{-1}$. The upper current of ASMMO increases with the increase of $H_c$; it increases from $I = 1.38 I_{th}$ when $H_c = 0.01$ m$^{-1}$s$^{-1}$ to $I = 1.65 I_{th}$ when $H_c = 0.12$ m$^{-1}$s$^{-1}$. The further increase of current beyond the region of ASMMO results in conversion of laser oscillation to SMO. On the other hand, the figures indicate that HMMO dominates the laser operation when the AGS parameter $H_c$ exceeds the upper boundaries of ASMMO and SMO and the current increase the boundary of $I \leq 1.12 I_{th}$ of the SMMO region.

In the following subsections, we present the characteristics of each of these types of modal oscillations in the time domain and the corresponding frequency properties of RIN.

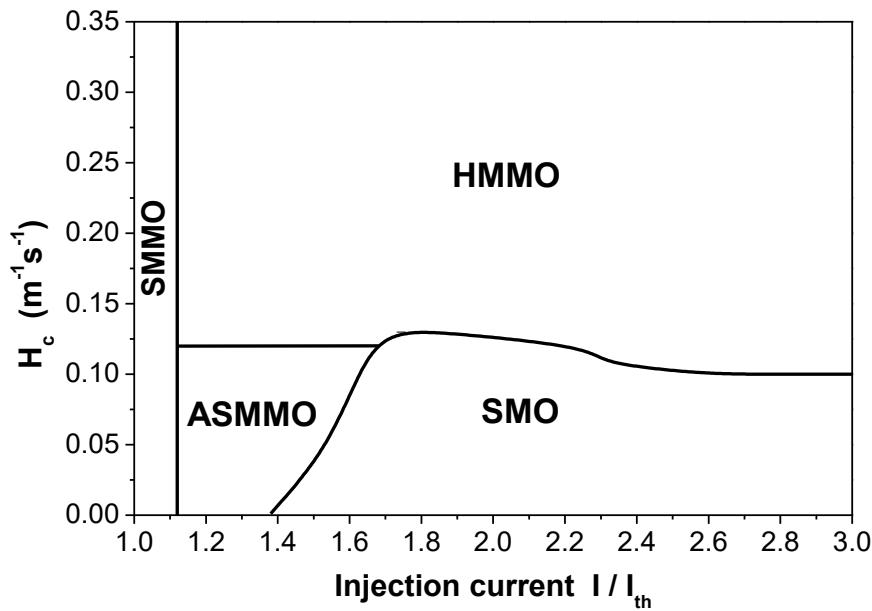

**Figure 1.** ($H_c$ versus $I$) diagram of modal oscillation. The four modal oscillation types are allocated.

## A. Hopping multimode oscillation (HMMO)

This type is dominating the dynamics of long-wavelength lasers, such as 1.33μm and 1.55μm-InGaAsP lasers because they have large values of the α-factor and differential gain $a$ [23,24] and hence the $H_c$ parameter. In this case AGS is strong enough to induce multimode hopping. Figure 2(a) plots typical temporal trajectories of the strongest three modal intensities $S_p(t)$ as well as that of the total laser intensity $S(t)$. These strongest modes lie on the longer-wavelength side of the central mode ($p = +2, +3, +4$) as a typical manifestation of AGS. As shown in the figure, these strongest modes do not reach steady-state operations; the dominant mode suddenly drops-off followed by an instantaneous jumping of one of the other modes. This hopping is repeated among the modes in a semi-periodic style, resulting in a semi-periodic switching of modes to the lasing level as $p = +2 \rightarrow +3 \rightarrow +4 \rightarrow +2$, and so on. The rotating frequency of this mode hopping is $f_{HM} = 67$ MHz. These simulation observations were interpreted basing on the Bogatov effect of AGS [25] that there are a number of modes lasing simultaneously, and the photons in each mode are scattered into the other modes causing either gain enhancement or suppression to these modes. Therefore, there exist more than one steady state, however only one of them is stable shifts to the long-wavelength mode by increasing the α-factor [25]. The intrinsic intensity fluctuations as fluctuations of the switching state of the hopping modes as well as fluctuations of the total intensity around the steady-state vale. These hopping modes share in the total laser output with SMSR = 1.6, rusting in the multimode spectrum of figure 2(b). This spectrum is asymmetric around the strongest mode of $p = +4$, which fits the recorded spectrum for 1.3μm-InGaAsP lasers by Yamada *et al.* [8]. The gain spectrum shown in figure 2(b) is also asymmetric and shallow with the maximum modal gain $G_{+4} < G_{th}$ and the gain difference being as small as $(G_{+4}-G_{+3})/G_{th} = 0.005$.

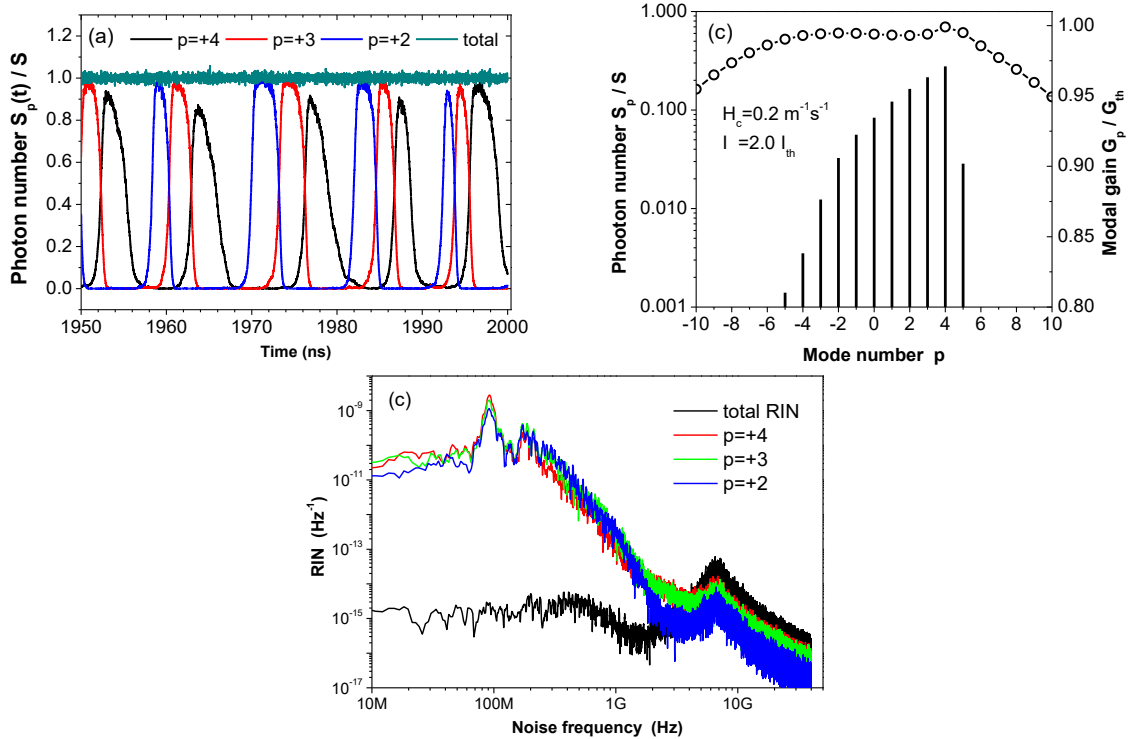

**Figure 2.** Typical characteristics of HMMO when $I = 2I_{th}$ and $H_c = 0.2$ m$^{-1}$s$^{-1}$ (strong AGS conditions): (a) time variation of modal intensity, and (b) output spectrum with modal gain spectrum on right axis, and (c) RIN spectra of the total output and hopping modes.

The corresponding frequency spectrum of RIN is given in figure 2(c). The spectrum exhibits a sharp peak around the RO frequency of the laser, $f_r$ = 7 GHz. The low-frequency part of RIN of the oscillating modes is much higher than that of the total RIN, a feature known as mode-partition noise [1]. This difference of noise is almost five orders of magnitude. The figure shows also that both the total RIN and RIN of the hopping modes exhibit the RO peak. The figure displays also the interesting multimode hopping peaks around 67 MHz, which corresponds to the frequency of semi-periodic switching of the hopping modes to the lasing level seen in figure 2(a). Such peaks were observed in the measured RIN spectrum of InGaAsP lasers [8]. In this case, the low-frequency RIN of the total power is LF-RIN = $2 \times 10^{-15}$ Hz$^{-1}$.

## B. Symmetric multimode oscillation (SMMO)

This type dominates the laser dynamics when the laser operates near the threshold level, $I <$ 1.1$I_{th}$, regardless the value of $H_c$. In this case, the laser output is mainly contained in a number of modes lying on both sides of central mode $p$ = 0. An example of the output mode spectrum and the associated gain spectrum are shown in figure 3(a) when $H_c$ = 0.12 m$^{-1}$s$^{-1}$. SMSR is as small as $S_0/S_{+1}$ = 8.1 and the laser is said to oscillate in multimode. The figure indicates a symmetric spectrum around the central mode $p$ = 0, which is the conventional multimode type that characterizes semiconductor lasers operating close to the threshold level [26]. The gain spectrum is nearly symmetric, homogeneously distributed around the gain center and shallow the gain difference is as small as $(G_{+1}-G_0)/G_{th}$ = 6.3x10$^{-4}$.

The frequency spectrum characterizing this type is plotted in figure 3(b). The low-frequency noise of the total output is flat with an average level of LF-RIN = 4.6x10$^{-14}$ Hz$^{-1}$, while those of the strongest mode is lower by a three-order of magnitude. That is the mode-partition noise is lower in this case than the case of HMMO. The peak of RO occurs around frequency $f_r$ = 1.3GHz, which is lower than that of MMHO because the laser is injected by a lower current of $I$ = *1.05I$_{th}$* which corresponds to stronger effect of the spontaneous emission in the lasing action.

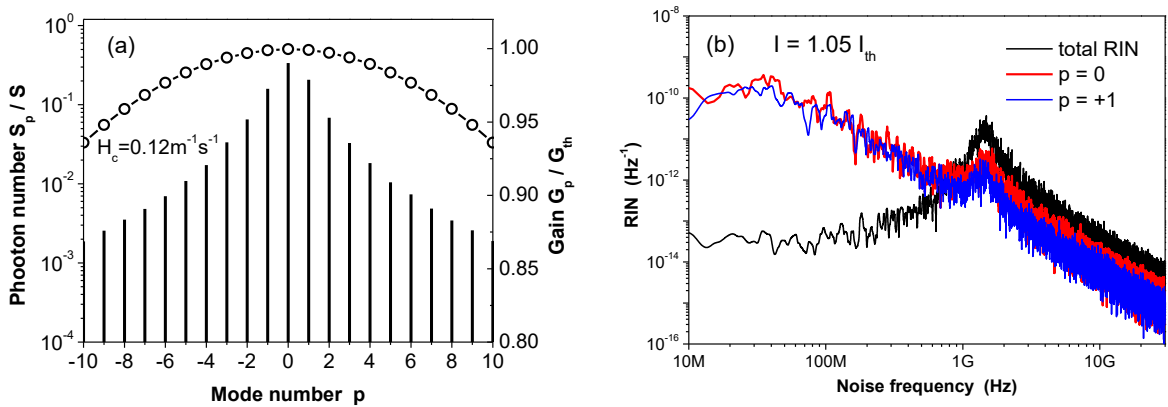

**Figure 3.** Typical characteristics of SMMO when $I$ = 1.05$I_{th}$ and $H_c$ = 0.12 m$^{-1}$s$^{-1}$: (a) output and modal gain spectra, and (b) RIN spectra of the total output and strongest modes.

## C. Asymmetric steady-state multimode hopping (ASMMO)

This is another type of multimode oscillations in which the laser output is contained in a number of long-wavelength modes. In this case, the oscillating modes compete in the transient region and then attain the steady state after few nanoseconds keeping SMSR < 100 [15]. The output mode spectrum and the gain spectrum are shown in figure 4(a), which corresponds to $I = 1.2I_{th}$ and $H_c = 0.12$ m$^{-1}$s$^{-1}$. As seen in figure 4(a), the output spectrum is asymmetric around the strongest mode $p = +1$, and the spectrum corresponds to multimode oscillations with SMSR = $S_{+1}/S_{+2}$ = 21.3. The asymmetric character of the modal oscillation is seen also in the spectral gain profile of figure 4(a) which is characterized also by shallow gain spectrum with $(G_{+1}-G_0)/G_{th} = 0.001$.

Figure 4(b) plots the frequency spectrum of RIN associated with ASMMO of figure 4(a). The difference of RIN of the total output with those of the strongest modes is five orders of magnitude like the case MMHO of figure 2(c). However, the RIN spectrum does not reveal the peak of mode hopping in the low-frequency regime of figure 2(c). Another difference is that peak of relaxation oscillations of mode $p = +2$ has a much lower level of RIN compared with the strongest mode $p = +1$.

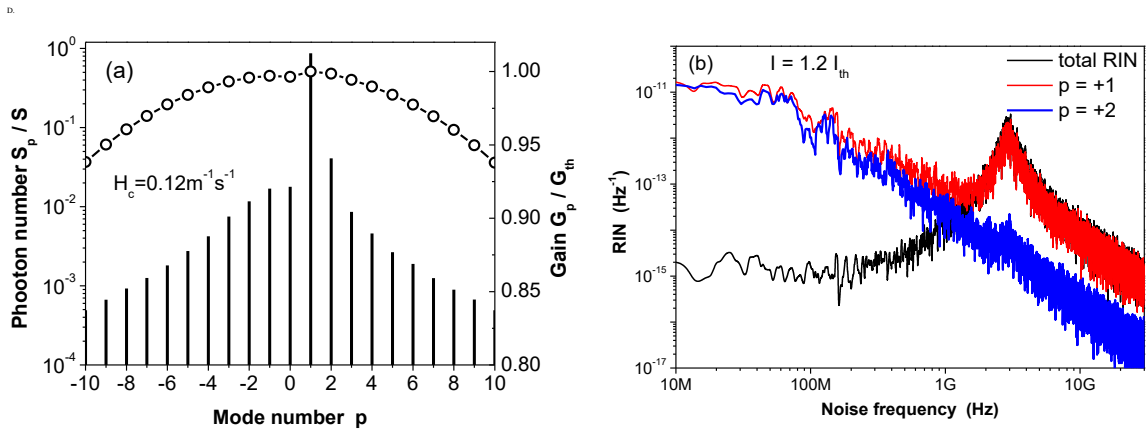

**Figure 4.** Typical characteristics of ASMMO when $I = 1.2I_{th}$ and $H_c = 0.12$ m$^{-1}$s$^{-1}$: (a) output and modal gain spectra, and (b) RIN spectra of the total output and hopping modes.

## E. Single mode oscillation (SMO)

This is the case of converting the multimode oscillations into oscillations in one mode. Figures 5(a) and (b) plot the output intensity and gain spectra of this type simulated when $H_c = 0.034$ and 0.12 m$^{-1}$s$^{-1}$, which correspond to SMSR = 713 and 224, respectively. The figure shows that the dominant mode jumps from $p = 0$ to $p = +3$ with the increase of the AGS which is associated also with degradation of SMSR. In these cases, the strongest mode achieves gain of unity and the gain differences between the dominant and strongest side mode are bigger than the cases of multimode oscillations; $(G_0-G_{+1})/G_{th} = 0.008$ and $(G_{+3}-G_{+4})/G_{th} = 0.002$ when $H_c = 0.034$ and 0.12 m$^{-1}$s$^{-1}$, respectively

The frequency spectrum of SMO when $H_c = 0.034$ m$^{-1}$s$^{-1}$ is plotted in figure 6. The low-frequency regime of the RIN spectrum is flat (white noise) for both the total output and strongest modes, with the RIN difference is almost two orders of magnitude. This noise is lowest among the investigated oscillation types with LF-RIN = $2.8 \times 10^{-17}$ Hz$^{-1}$. In the high-frequency regime, both RIN spectra of the total output and dominant mode reveal the RO peak around at frequency of $f = 6.7$GHz, which is lower than the peak frequency of figure 2(b) of MMHO since the current is lower in this case $I = 1.8I_{th}$. On the other hand RIN of the strongest side mode $p = +1$ drops to lower magnitudes with the increase of the noise frequency $f$.

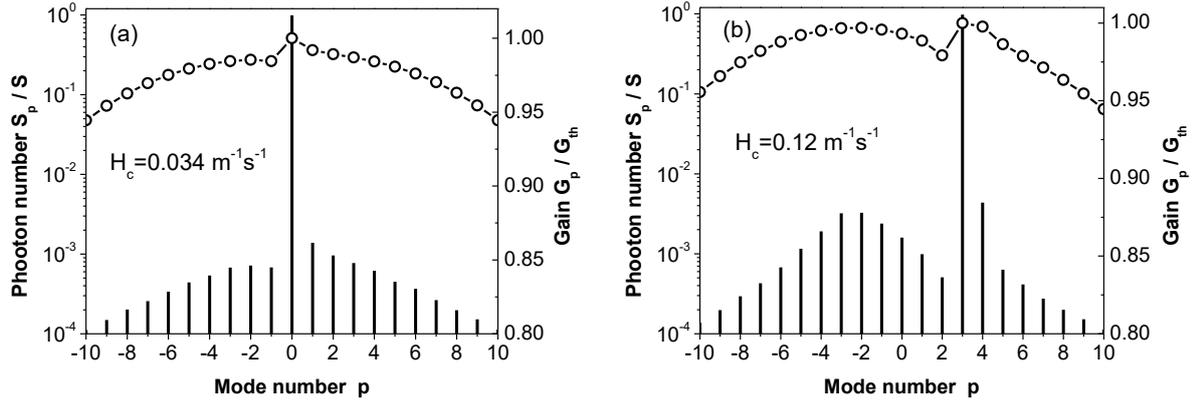

**Figure 5.** Output spectrum and modal gain under SMO when (a) $H_c = 0.034$ m$^{-1}$s$^{-1}$ and (b) $H_c = 0.12$ m$^{-1}$s$^{-1}$ with $I = 1.8I_{th}$.

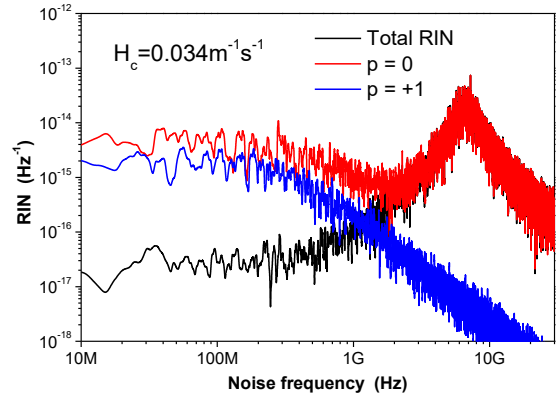

**Figure 6.** Typical RIN spectrum of the total output and predominant modes of SMO when $H_c = 0.034$ m$^{-1}$s$^{-1}$ and $I = 1.8I_{th}$.

*4.2 Influence of asymmetric gain parameters*

In this subsection we investigate influence of AGS parameter $H_c$ on the noise levels of the total output and the strongest mode of the investigate laser. Figure 7(a) and (b) plot variation of LF-RIN with $H_c$ when the laser is biased by currents of $I = 1.8I_{th}$ and $3.0I_{th}$. Both figures correspond to the cases of SMO and HMMO, as investigated in figure 1. The figures indicate that the RIN levels are lower when the laser exhibits SMO than the case of multimode oscillations. In the regimes of SMO, the RIN levels are lower in figure 7(b) than in figure 7(a) because of the reduction of the bias current. Also, the RIN levels increase with the increase of $H_c$, which is associated with degradation of the SMSR. When the SMSR becomes less than 100 and the laser exhibits HMMO, RIN abruptly increases. The figures show also a decrease in the RIN difference between the total laser output and the strongest mode from figure 7(a) to figure 7(b), which indicates reduction of the mode partition noise.

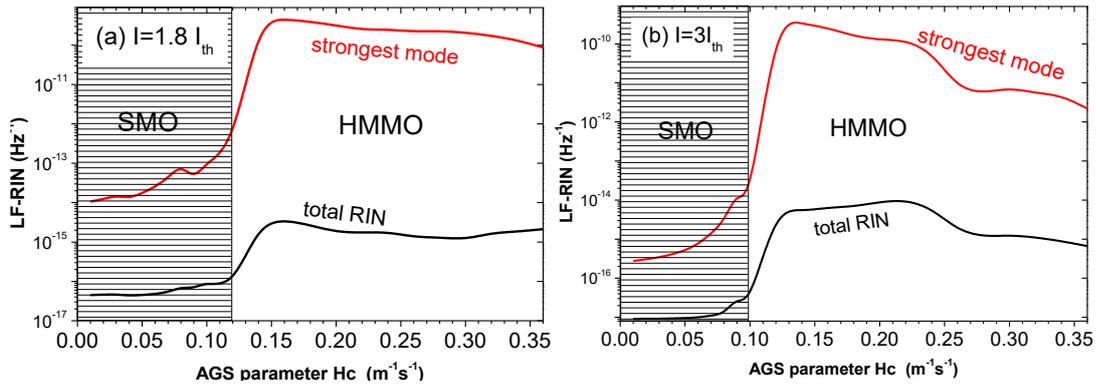

**Figure 7.** Variation of LF-RIN with the ASG parameter $H_c$ when (a) $I = 1.8I_{th}$, and (b) $I = 3.0I_{th}$.

*4.3. Influence of injection current I*

Influence of the bias current $I$ on the noise properties of the laser is illustrated in figure 8. The figure plots variation of LF-RIN with $I$ when $H_c = 0.12$ m$^{-1}$s$^{-1}$. In this case and as figure 1 indicates, the laser oscillates in SMMO near the threshold level, and with the increase of $I$ the laser converts into ASMMO followed by oscillations in SMO before entering the operating region of HMMO. The figure shows that except for the region of HMMO, LF-RIN decreases with the increase of current $I$ which reflects improvement of laser coherency [7]. In the region of HMMO, both LF-RIN and the difference in RIN between the total output and strongest mode increase with the increase of $I$ which is associated with reduction of the SMSR. The RIN level is lowest in the region of SMO and highest when the laser oscillates in SMMO (because of the spontaneous emission noise) and in the most noisy region of HMMO.

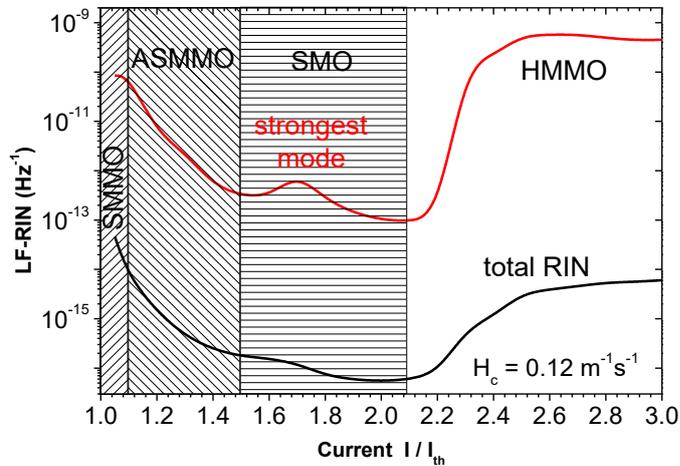

**Figure 8.** Variation of LF-RIN with current $I$ when the ASG parameter is $H_c = 0.12$m$^{-1}$s$^{-1}$.

Before we conclude, we will offer some insights into how the above findings can be augmented to the regime of miniaturized cavities [27-29,39-44] and nanoscale laser physics [30-37]. Indeed, a key question debated yet until today is whether, or not, the effects of ASG and can be influenced by the Purcell factor, $F_p$, which captures the light-matter-interaction strength such as of a laser cavity and is proportional to the cold-cavities' quality factor (Q) divided by the cavities' mode volume [30]. The Purcell factor is especially pronounced for nanoscale light emitters and lasers, primarily due to the nonlinear scaling of volume and introduced loss due to the cavities's inability to provide feedback. This is especially of relevance for SMO in these compact volume cavities at (or below) the diffraction-limit of light [30-33]. Indeed, it was argued that laser designs with enhanced $F_p$ are also capable of also increasing the temporal RO of the laser cavity, thus expanding the 'speed' of the laser under direct modulation. Beyond semiconductor lasers, it would be interesting to explore ASG further also in cavities with high longitudinal modes such as present in optical fiber laser systems for Brillouin amplification and lasing, for instance [38].

## 5. Conclusions

We characterize the frequency spectrum of RIN in multimode semiconductor laser with enhanced AGS that induces multimode hopping. The possible types of modal oscillations were classified by scaling an AGS parameter $H_c$, which is determined by the linewidth enhancement α-factor and differential gain coefficient. These two parameters can be controlled by scheming the amplifying layer of the laser. We presented a mapping ($H_c$ versus $I$) diagram of the modal oscillation by intensive numerical integrations of the stochastic rate equations of the laser. Basing on the results obtained in the present study, conclusions can be itemized as follows:

(1) In the regime of small values of $H_c < 0.12$ m$^{-1}$s$^{-1}$ (weak AGS), the increase of current induces the types of modal oscillation of SMMO → ASMMO → SMO. In the regime of large values of $H_c$ (strong), SMO converts directly to HMMO.
(2) The frequency spectrum of the total RIN and the hopping-mode RIN of HMMO are characterized by a sharp peak around the RO frequency as well as a broad peak around the hopping frequency. On the other hand, under SMO, these RIN spectra are flat in the regime of low frequencies, and only the total RIN and that of the dominant mode exhibit the RO peak.
(3) The levels of RIN in the regimes of SMO are much lower than those in the noisiest region of HMMO. The difference between the total RIN and the strongest mode RIN, which is a measure of MPN, is larger than four orders of magnitude but is reduced to two orders of magnitude when the oscillation is converted to SMO.
(4) Contrary to the region of SMO in which LF-RIN, LF-RIN increases with the increase of $I$ due to enhancement of AGS and reduction of SMSR

**Acknowledgment**

This project was funded by the Deanship of Scientific Research (DSR) at King Abdulaziz University, Jeddah, under grant no. (**RG-18-130-41**). The authors, therefore, acknowledge with thanks DSR technical and financial support.